%Paper: hep-ph/9209229
%From: FADAMS@mail.physics.lsa.umich.edu
%Date: Fri, 11 Sep 92 17:18 EDT

%--- Version with plane parallel treatment -- 18 August, 1992
\magnification=1200
\hoffset=0.1truein
\voffset=0.1truein
\vsize=23.0truecm
\hsize=16.25truecm
\parskip=0.2truecm
\def\newpage{\vfill\eject}

\def\vbub{ v_0 }
\def\ninf{ n_\infty }
\def\Tinf{ T_\infty }
\def\lt{ T_\Delta }
\def\dwig{ {\widetilde d} }

\def\Rnuc{ R_{nuc} }
%\baselineskip=24pt
%
%---------------------------------------------------------------------
%
$\,$
\centerline{}
\centerline{\bf BARYON NUMBER DIFFUSION AND INSTABILITIES}
\centerline{\bf IN THE QUARK/HADRON PHASE TRANSITION}
\vskip 0.2truein
\centerline{\bf Fred C. Adams,$^{1,2}$ Katherine Freese,$^{1,2}$
and J. S. Langer$^2$ }
\vskip 0.3truein
\centerline{\it $^1$Physics Department, University of Michigan}
\centerline{\it Ann Arbor, MI 48109}
\vskip 0.15truein
\centerline{\it $^2$Institute for Theoretical Physics}
\centerline{\it University of California, Santa Barbara, CA 93106}
\vskip 0.4truein

\centerline{\it submitted to Physical Review Letters}
\centerline{\it 26 August 1992}

\bigskip

\medskip
\bigskip
\centerline{\bf Abstract}
\medskip

Hadron bubbles that nucleate with radius $R_{nuc}$
in a quark sea (if the phase transition is first
order) are shown to be unstable to the growth of
nonspherical structure when the bubble radii
exceed a critical size of $20 - 10^3$ $R_{nuc}$.
This instability is driven by a very thin layer
of slowly diffusing excess baryon number that
forms on the surface of the bubble wall and
limits its growth. This instability occurs on a
shorter length scale than those studied previously
and these effects can thus be important for both
cosmology and heavy ion collisions.


\newpage

If a first order phase transition occurs when quarks are confined into
hadrons at a temperature $T_C$ = 100 -- 200 MeV, many interesting
physical consequences can arise [1].  Previous investigations [1, 2]
have shown that the baryon number density can be higher in the quark phase
than in the hadron phase. In the case of cosmology, this asymmetry can
lead to inhomogeneous distributions of baryons after the phase transition
[1, 3]. This possibility has led to numerous studies of big bang
nucleosynthesis in the presence of baryon inhomogeneities [3 -- 7].
Another possible relic of such a phase transition is stable macroscopic
quark matter -- quark nuggets -- which may survive to the present epoch
and contribute to the dark matter of the universe [1, 8].  The effects
of a first order quark/hadron (Q/H) phase transition are also,
in principle, detectable in relativistic heavy ion collisions.

In this Letter, we assume that the Q/H phase transition is first
order and study a new instability in the growth of hadron bubbles.
When the background plasma, either in the early universe or in a
heavy ion collision, cools to a temperature below the critical
temperature $T_C$, bubbles of the hadronic phase nucleate and begin
growing spherically. Witten [1] has argued that baryon number prefers
to be in the quark phase.  As a result, the bubble rejects baryon
number as it grows.  Because baryon number diffuses very slowly, it
is confined and concentrated into a thin layer on the surface of the
bubble.  We argue that this boundary layer must limit the growth of
the bubble and cause it to become morphologically unstable.

In this analysis we assume that the latent heat of the phase
transition is transported radiatively, mostly by neutrinos,
since they have the longest mean free path. The relative
importance of radiation and hydrodynamic flow in removing the
latent heat remains unsettled [1, 4, 9, 10].  Previous work has
studied the (stable) evolution of hadron bubbles when hydrodynamic
flow dominates the removal of latent heat [11, 12]. However, radiative
processes are present and perhaps dominant from the moment of bubble
nucleation onward.  In addition, we expect that the new instabilities
(due to baryon number diffusion) presented here can arise regardless
of the type of heat flow.

In previous work [9], we showed that when radiative effects dominate,
hadron bubbles are unstable for bubble radii greater than the mean free
path of the neutrino ($\sim$10 cm) because the heat flow becomes diffusive
at that size scale.  Hydrodynamic instabilities during hadron bubble growth
[13], if present, occur on a smaller scale of order $10^{-6}$ cm.  The
instability proposed here occurs on yet smaller scales, at most
$10^5$  fm, and thus greatly alters the manner in which the phase
transition proceeds.  The resulting inhomogeneities in baryon number
are too small (in length scale) to affect big bang nucleosynthesis.
In fact, this size scale is small enough that this new instability can
occur in heavy ion collisions, whereas the previously studied instabilities
cannot. We emphasize, however, that significant uncertainties exist in both
the QCD physics and our knowledge of nonequilibrium pattern formation.

We start by considering a locally flat section of a bubble wall
at which the hadronic phase is advancing at speed $v$ into the
quark sea. Previous work [1, 2] has shown that, when these phases
are in equilibrium with each other, the baryon number in the quark
phase $n_q$ can be much higher than that in the hadron phase $n_h$;
the ratio $n_h/n_q$ is typically 0.003 -- 0.2.  For simplicity, we
assume that this ratio vanishes and henceforth consider only
quantities in the quark phase.

In a frame of reference moving with the bubble wall, the diffusion
equation for the baryon number density $n$ in the quark phase is
$$ {\partial n \over \partial t} - v {\partial n \over \partial z}
= D \nabla^2 n , \eqno(1)$$
where the direction of motion is along the $z$-axis and
where $D$ is the diffusion constant.  We will assume that
$v$ $\approx$ {\it constant}; we justify this assumption
below. Since quarks carry the baryon number in this phase,
the diffusion constant is given by
$D = \langle v_q \rangle \ell_q$/3,  where $\langle v_q \rangle$
is the mean speed of the quarks ($\langle v_q \rangle$ is fairly close
to the speed of light $c$) and $\ell_q$ is the mean free path of
a quark (about 1 fm for $T_C$ $\sim$ 200 MeV).

Deep in the quark phase, very far from the advancing hadronic front,
the baryon number density has a value appropriate for thermal
equilibrium at temperature $\Tinf$, i.e., $\ninf = g$
$(\mu/T)_\infty \, \Tinf^3$, where the constant $g$ depends on
the number of degrees of freedom in the quark phase and where $\mu$
is the chemical potential. For our universe, standard big bang
nucleosynthesis [14, 15] constrains $\mu$ to be very small, i.e.,
$(\mu/T)_\infty$ $\sim$ $10^{-9} - 10^{-8}$.  For heavy ion
collisions, $(\mu/T)_\infty$ $\sim$ $10^{-4} - 10^{-2}$.

The diffusion equation (1) must be supplemented by boundary conditions
at the Q/H interface.  We assume local thermodynamic equilibrium, so
that the baryon number density $n_I$ and temperature $T_I$ at the
interface are related by a Gibbs-Thompson condition [16, 17] which,
in this case, takes the form
$$T_I = T_C (1 - d_0 \kappa) - \beta \lt n_I^2 . \eqno(2)$$
Here, $\kappa$ is the interfacial curvature;
$d_0$ $\equiv \sigma T_C C_p / L^2$ is a capillary length and
is proportional to the surface tension $\sigma$;
the specific heat is $C_p$ = ${\cal O}(T_C^3)$;
and the latent heat is $L$ = ${\cal O}(T_C^4)$.
Lattice gauge theory calculations [18] have estimated the surface
tension to be $\sigma$ $\approx$ 0.1 $T_C^3$.  The third term in Eq. (2),
$\beta \lt n_I^2$, is the decrease in coexistence temperature due to
the presence of baryon number (solute).  Notice that this term is
quadratic rather than linear in the solute concentration $n_I$.
We have defined a convenient energy scale $\lt \equiv {L / C_p}$,
which we expect to be ${\cal O}(T_C)$.
To estimate the value of $\beta$, we use a bag model for the equation
of state (and we take $n_h/n_q$ = 0).  We find that $\beta$ $\approx$
0.6 $(T_C/\lt) T_C^{-6}$ $\approx$ 0.6 $T_C^{-6}$, where we have
used the results of Ref. [11] in obtaining the numerical
coefficient.

Continuity of baryon number at the interface requires
$$ n_I v_I = - D \Bigl[ {\hat \nu} \cdot \nabla n \Bigr]_{z=z_I} ,
\eqno(3)$$
where $z_I$ is the interface position, $\hat \nu$ is the unit normal
to the interface, and $v_I$ is its speed in the $\hat \nu$-direction.
The physical meaning of Eq. (3) is that the rate at which excess
baryon number is rejected from the interior of the hadronic phase is
balanced by the rate at which baryon number diffuses ahead of the
two-phase interface. This effect produces the destabilizing boundary
layer of excess baryon number on the surface of the growing bubble.

To complete the statement of our model, we must impose a condition
of heat continuity at the moving interface.  We can consider a full
treatment of the heat transport either through a diffusion equation
for temperature or through a radiative transfer equation [9].  For
the parameter regime of interest, however, neutrinos carry away most
of the energy and the mean free path of the neutrino is $\sim$ 10 cm,
much greater than the size scale of the bubble.  Hence, the condition
of thermal continuity reduces to the simple form [9]
$$\vbub = c {T_I - \Tinf \over \lt} . \eqno(4)$$
Our assumption here is that all of the latent heat produced in the
phase transition is efficiently carried out of the system by radiation
and that the interface moves at the fastest speed for which this
condition holds.  [We have implicitly assumed that the bubbles are far
enough apart that the latent heat released during the evolution of a
given bubble does not affect other bubbles.  If, however, the mean
bubble separation is less than the neutrino mean free path, mutual
heating effects should be considered.]

We now consider the stability of the moving bubble wall.
As we will see, the situations of interest to us are those for which
$n_I \gg \ninf$.  In this case, the baryon number density in the
boundary layer ahead of the unperturbed wall at $z_I = 0$ is
accurately given by Eqs. (1) and (3) to be $n_I \exp(-2 z /\ell)$,
where the layer thickness $\ell = 2 D/v$ is very small for small
$D$. To study the stability of this wall against periodic
deformations of wave-vector $\bf k$, we consider linear
perturbations of the form
$$\delta n(z, {\bf x}, t) \sim {\hat n}_k \exp \bigl[
i {\bf k} \cdot {\bf x} - q_k z + \omega_k t \bigr] \, , $$
$$\delta z_I ({\bf x}, t) \sim {\hat z}_k \exp \bigl[
i {\bf k} \cdot {\bf x} + \omega_k t \bigr] \, , \eqno(5)$$
$$\delta T_I ({\bf x}, t) \sim {\hat T}_k \exp \bigl[
i {\bf k} \cdot {\bf x} + \omega_k t \bigr] \, , $$
where $\bf x$ denotes the position in the plane of the wall.
After performing a standard stability analysis [19], we find
that the amplification rate $\omega_k$ is
$$\omega_k = { v \, (q_k \ell - 2) / \ell \over
1 + {D (q_k \ell - 2) / 2 \beta n_I^2 c \ell} }
\Biggl[ 1 - {\dwig \ell k^2 \over 4 \beta n_I^2}
\Biggr] \, , \eqno(6)$$
where $\dwig$ $\equiv$ $(T_C/\lt) d_0$ and $q_k \ell$ $\equiv$
1 $+ (1 + k^2 \ell^2 + \omega_k \ell^2/D)^{1/2}$.
The crucial result is that the wall is unstable, i.e.,
$\omega_k$ $>$ 0, for wave-vectors $k < k_C$,
where   $$k_C \equiv \bigl( {4 \beta n_I^2 / \dwig \ell}
\bigr)^{1/2} \, . \eqno(7)$$

We now propose two separate estimates of the characteristic
length scales for patterns produced by this instability.
First, consider a growing hadronic bubble with radius $R$
and $v = {\dot R}$.  If we assume that all the baryon number
that was swept out by the growing bubble remains in the boundary
layer at its surface, then $2 \pi R^2 \ell n_I$ =
$4 \pi R^3 \ninf /3$.  With this assumption, Eqs. (2) and (4)
can be combined to yield a differential equation for $R(t)$:
$${v \over c} = { {\dot R} \over c} = \Delta
- {2 \dwig \over R} - \beta \ninf^2 \Bigl( {R {\dot R} \over 3 D}
\Bigl)^2 , \eqno(8)$$
where $\Delta \equiv (T_C - \Tinf)/\lt$ is a dimensionless measure
of the undercooling. The condition $\dot R$ = 0 determines the
nucleation radius $R_{nuc} = 2 \dwig / \Delta$.  For $T_C$ = 200 MeV
and $\sigma = 0.1 T_C^3$, $\Rnuc$ $\approx$ (0.2 fm)/$\Delta$.
The relevant value of $\Delta$ is unknown, although cosmological
studies usually assume small values, e.g., $\Delta$ $\sim$
$10^{-2}$ -- $10^{-4}$; for heavy ion collisions, we assume a
larger value $\Delta \sim$ 1. The bubble becomes unstable for
$R \gg R_{nuc}$.  When the instability occurs, however, the third
term in Eq. (8) is still very small (the results presented below
verify this claim); thus, our implicit assumption of a nearly
constant growth speed $v$ is valid.

Our stability analysis implies that the bubble becomes
morphologically unstable when its radius is appreciably
larger than $R_C$ = $2 \pi/ k_C$.  (This result can easily
be confirmed by a more systematic analysis in spherical
coordinates.) Using our formula for $k_C$ and the conservation
of baryon number assumption to eliminate $n_I$, we find
$${ R_C \over \Rnuc } \approx \Biggl[ {D \over c \dwig} \Biggr]^{3/4}
\, \, \Biggl[ \, {\Delta \over \beta \ninf^2} \, \Biggr]^{1/4}
\,  . \eqno(9)$$
Notice that $D/c\dwig$ is of order unity. For cosmology,
$\beta \ninf^2$ $\sim$ $10^{-17}$. If we take an undercooling
of $\Delta = 10^{-3}$, we find $R_C / R_{nuc}$ $\sim$ $10^3$, or
$R_C$ $\sim$ $10^5$ fm.  [Notice that our approximation
$n_I/\ninf \gg$ 1 is valid; for this case, $n_I/\ninf$
= $2 R/3\ell$ $\approx$ $10^3$.]  For heavy ion collisions,
we obtain $R_C / R_{nuc}$ $\sim$ 20, or $R_C \sim$ 4 fm.

The problem with the above argument is that it says little
about the final configuration of the bubble. Indeed, linear
stability analysis is notoriously unreliable for predicting
quantitatively the final relevant length scales for
such highly nonlinear processes.  Unfortunately,
the present situation is not, as far as we know, directly
analogous to pattern-forming systems that have been studied
experimentally. The closest analogy that comes to mind is
directional solidification of a dilute alloy where a planar
front becomes unstable and breaks up into elongated cellular
structures.  The excess solute (here, baryon number) that
accumulates ahead of the front diffuses along the boundary
layer and ultimately is trapped, in concentrated form,
in the narrow, sometimes filamentary, interstices between
the cells.  The filaments themselves are usually unstable
and break up into chains of droplets, which are made up of
concentrated solute and trail behind the moving front.
Although this mechanism is well known in metallurgy, no
reliable theory for predicting the cellular spacing is available.

As our second estimate of characteristic length scales produced
by the instability, we propose the following.  Suppose that the
Q/H interface leaves
behind it baryon-rich filaments with characteristic spacing $W$.
Clearly, $W$ cannot be larger than the stability length
$2 \pi / k_C$; we therefore assume marginal stability and set
$W$ $\sim$ $k_C^{-1}$.  Let us further assume that the radii of
the filamentary interstices, where they join the moving interface,
are roughly equal to the diffusion length $\ell$ or, equivalently,
$R_{nuc}$, both of which are about the same size and seem to be
the only relevant length scales. (We expect that this assumption
is the weakest part of our argument.  In any case, further work
should be done to check its validity.) We can now use global
conservation of baryon number, i.e., $\ninf W^2 \approx n_I \ell^2$.
{}From this result and our formula for $k_C$, we find
$${W \over R_{nuc} } \sim \Biggl[ {\Delta \over \beta \ninf^2}
\Biggr]^{1/6} \, , \eqno(10)$$
which implies cosmological values of $W$ about one order of
magnitude smaller than those predicted by Eq. (9).

We can take this argument one step further and estimate the size
of the filaments far behind the advancing front.  The filaments
will contract to a radius $\varpi$ where the baryon number
density $n_f$ is sufficiently large that $T = \Tinf$ and
$v = 0$; the Gibbs Thompson relation applied on a tubular
surface is
$$\Delta + \dwig / \varpi = \beta n_f^2 , \eqno(11)$$
where the curvature has changed sign (from Eq. [2]) because the
hadronic phase is now on the outside. Conservation of baryon number
requires that $n_I \ell^2 \approx n_f \varpi^2$, and we thus obtain
$$ {\varpi \over R_{nuc} } \sim \Biggl[ { \beta \ninf^2 \over \Delta}
\Biggr]^{1/9} \, , \eqno(12)$$
which implies that $\varpi$ is of order 1 fm. Thus, the size scale of
any baryon number inhomogeneities produced in the phase transition
will be much smaller than that required ($>$ 100 cm) to affect
big bang nucleosynthesis.

\medskip

In summary, we have shown that the growth of hadron bubbles
during the Q/H phase transition can be unstable due to a new
instability mechanism. The instability first sets in when
the bubble grows to a radius of $\sim 20 - 10^3$ times the original
nucleation radius. These instabilities can greatly alter the long term
evolution of the bubbles. We have presented here a marginal stability
hypothesis in which the moving interface leaves behind tubular regions
which are rich in baryon number and which ultimately break up into
droplets of size $\varpi \sim$ 1 fm.

The most important of our assumptions are as follows:
(A) We have assumed that the Q/H phase transition is in fact first
order. We stress that the evidence on this issue remains divided.
(B) We have assumed local thermodynamic equilibrium and have written the
thermodynamic condition at the bubble interface as the Gibbs Thompson
relation given in Eq. (3). The true form of this relation depends on
the details of QCD physics which are not yet well understood.
(C) We have assumed that hadron bubble growth is limited by the
diffusion of baryon number away from the interface. This assumption
is expected to be valid for sufficiently large baryon number densities
(or, equivalently, large $\mu/T$) and/or sufficiently large bubble
radii. We note that the first of these conditions is more likely to
be met in heavy ion collisions, whereas the second is more likely to
be met in the early universe.
(D) We have assumed that all of the latent heat produced by the phase
trasition is carried away by radiation and hence no hydrodynamic flow
occurs.  We note that even when this assumption is violated, the
requirement of baryon number diffusion can still lead to instabilities
which are qualitatively similar to those presented in this paper.
(E) We have invoked a marginal stability hypothesis to describe the
long term evolution of the system. This type of behavior is not well
understood even in laboratory systems, much less in the early universe.

\bigskip
\centerline{\bf Acknowledgements}
\medskip

We would like to thank D. Kessler, L. Sander, and T. Matsui for
helpful conversations and the ITP at U.C. Santa Barbara, where the
inspiration for this work took place, for hospitality.
FCA is supported by NASA Grant No. NAGW--2802.
KF is supported by NSF Grant No. NSF-PHY-92-96020, a Sloan
Foundation fellowship, and a Presidential Young Investigator award.
JSL is supported in part by NSF Grant No. NSF-PHY-89-04035 and
DOE Grant No. DE-FG03-84ER45108.

\newpage
\bigskip
\centerline{\bf References}
\medskip

\item{[1]} E. Witten, {\it Phys. Rev.} {\bf D}
{\bf 30}, 272 (1984).

\item{[2]} J. I. Kapusta and K. A. Olive, {\it Phys. Lett.}
{\bf B} {\bf 209}, 295 (1988).

\item{[3]} J. H. Applegate, C. J. Hogan, and R. J. Scherrer,
{\it Ap. J.} {\bf 329}, 572 (1988).

\item{[4]} J. H. Applegate and C. J. Hogan,
{\it Phys. Rev.} {\bf D} {\bf 31}, 3037 (1985).

\item{[5]} G. M. Fuller, G. J. Mathews, and C. R. Alcock,
{\it Phys. Rev.} {\bf D} {\bf 37}, 1380 (1988).

\item{[6]} C. Alcock, G. M. Fuller, and G. J. Mathews
{\it Ap. J.} {\bf 320}, 439 (1987).

\item{[7]} R. A. Malaney and W. A. Fowler,
{\it Ap. J.} {\bf 333}, 14 (1988).

\item{[8]} E. Farhi and R. L. Jaffe, {\it Phys. Rev.} {\bf D}
{\bf 30}, 2379 (1984).

\item{[9]} K. Freese and F. C. Adams,
{\it Phys. Rev.} {\bf D} {\bf 41}, 2449 (1990).

\item{[10]} J. C. Miller and O. Pantano,
{\it Phys. Rev.} {\bf D} {\bf 40}, 1789 (1989).

\item{[11]} K. Kajantie and H. Kurki-Suonio, {\it Phys. Rev.}
{\bf D} {\bf 34}, 1719 (1986).

\item{[12]} M. Gyulassy, K. Kajantie, H. Kurki-Suonio, and
L. McLerran {\it Nucl. Phys.} {\bf B} {\bf 237}, 477 (1984);
H. Kurki-Suonio {\it Nucl. Phys.} {\bf B} {\bf 255}, 231 (1985).

\item{[13]} B. Link, {\it Phys. Rev. Lett.} {\bf 20}, 2425 (1992).

\item{[14]} J. Yang, M. S. Turner, G. Steigman, D. N. Schramm,
and K. A. Olive, {\it Ap. J.} {\bf 281}, 493 (1984).

\item{[15]} L. Kawano, D. N. Schramm, and G. Steigman,
{\it Ap. J.} {\bf 327}, 750 (1988).

\item{[16]} J. S. Langer, {\it Rev. Mod. Phys.} {\bf 52}, 1 (1980).

\item{[17]} D. A. Kessler, J. Koplik, and H. Levine,
{\it Advances in Physics} {\bf 37}, 255.

\item{[18]} S. Huang, J. Potvin, C. Rebbi, and S. Sanielevici,
{\it Phys. Rev.} {\bf D 42}, 2864 (1990); K. Kajantie,
L. K\"arkk\"ainen, and K. Rummukainen, {\it Nuc. Phys.}
{\bf B 357}, 693 (1991);  R. Brower, S. Huang, J. Potvin,
and C. Rebbi, BUHEP-92-3 (1992).

\item{[19]} This calculation is a generalization of the original
given in W. W. Mullins and R. F. Sekerka, {\it J. Appl. Phys.},
{\bf 34}, 323 (1963);  W. W. Mullins and R. F. Sekerka,
{\it J. Appl. Phys.}, {\bf 35}, 444 (1964).

\bye